\documentclass[preprint,aps]{revtex4}%
\usepackage{amsmath}
\usepackage{amsfonts}
\usepackage{amssymb}
\usepackage{graphicx}%

\begin{document}
\title{On the asymptotic methods for nuclear collective models}
\author{A. C. Gheorghe$^{b)}$, A. A. Raduta$^{a),b)}$}

\affiliation {$^{a)}$Department of Theoretical Physics and Mathematics,Bucharest
University, POBox MG11, Romania}
\affiliation {$^{b)}$Institute of Physics and Nuclear Engineering,
Bucharest, POBox MG6, Romania}

\begin{abstract}
Contractions of orthogonal groups to Euclidean groups are applied to analytic
descriptions of nuclear quantum phase transitions. The semiclassical
asymptotic of multipole collective Hamiltonians are also investigated.
\end{abstract}
\maketitle

\section{Introduction}
\label{sec:level1}
\renewcommand{\theequation}{1.\arabic{equation}}
\setcounter{equation}{0}

The notion of Lie algebra and Lie group contractions was first introduced by
Segal \cite{Seg} in 1951 and E. E. In\"{o}n\"{u} and E. P. Wigner \cite{IW} in
1953. The idea of contractions of Lie groups historically appeared in relation
with the non-relativistic limit $c\rightarrow\infty$, where $c$ is the speed of
light. That limit brings the Poincar\'{e} group of relativistic mechanics to
the Galilei group of classical mechanics. Then the Euclidean group $E(2)$ is
given by the contractions of the groups $SO(3)$ and $SO(1,2)$. In\"{o}n\"{u}
and Wigner have been interested in unitary representations of $E(2)$ in which
the spherical harmonics become Bessel functions. The other example is a limit
process from quantum mechanics to classical mechanics in the limit
$\hbar\rightarrow0$, where $\hbar$ is the Planck constant. This limit
corresponds to the contraction of the Heisenberg algebras to the Abelian ones
of the same dimensions. The dynamical symmetries are realized by Lie groups
and algebras, superalgebras, quantum groups and algebras. The essential idea
of singular and degenerate transformations is presented in all cases of
dynamical symmetry contractions. They lead to asymptotic relations between
basis functions for the representations of different groups.

In section 2, we study the contractions from the orthogonal group $SO(n+1)$ to
the Euclidean group $E(n)$. Moreover, we study the contractions of the
corresponding homogeneous spaces: the sphere $S_{n}\sim O(n+1)/O(n)$ contracts
to the Euclidean space $\mathbb{R}^{n}\sim E(n)/SO(n)$ in the limit
$R\rightarrow\infty$, where $R$ is the radius of $S_{n}$. It is well known
that representations of the Euclidean groups can be constructed in terms of
the Bessel functions \cite{Vil}. We will characterize the contractions from
$SO(n+1)$ to $E(n)$ in terms of asymptotic expressions such that the Bessel
functions are obtained from the Laguerre functions. We approximate the zeros
of the Bessel functions using zeros of Laguerre functions. We obtained this
approximation in \cite{RGF} for the nuclear model $E(5)$ \cite{Yac}. \ We
apply the results to the dynamical group contractions in the critical points
of nuclear phase transition for the models $E(2n+1)$ and $X(2n+1)$.

In section 3, we study the semiclassical spectral properties of collective
Hamiltonians. We consider the classical quadrupole collective Hamiltonian
$\mathcal{H}_{n}$ with polynomial potential of degree $n$. The Henon - Heiles
$\mathcal{H}_{3}$ is a non-integrable Hamiltonian. Using the
Birkhoff-Gustavson normal form method for $\mathcal{H}_{4}$, we explicitly
obtain an integrable form as function of two action variables and the
corresponding semiclassical quantum spectrum. Moreover, we consider a
spherical multipole quantum Hamiltonian and apply the logarithmic perturbation
theory for the radial Schr\"{o}dinger equation. We obtain the explicit energy
spectrum up to the order $\hbar^{5}$.

\section{Contractions of orthogonal groups \newline to Euclidean groups}
\label{sec:level2}
\renewcommand{\theequation}{2.\arabic{equation}}
\setcounter{equation}{0}
In this section we focus on analytic contractions of orthogonal groups to
Euclidean groups.

Let $V$ be a vector space over $\mathbb{R}$ or $\mathbb{C}$ and $f\colon
(0,1]\rightarrow GL(V)$ be a continuous function. Let $[,]$ be a Lie bracket
on $V$. A parametrized family of Lie brackets on $V$ is defined by
$[x,y]_{\varepsilon}=f_{\varepsilon}([f_{\varepsilon}^{-1}(x),f_{\varepsilon
}^{-1}(y)])$. If the limit $[x,y]_{0}=\lim_{\varepsilon\rightarrow
0}[x,y]_{\varepsilon}$ exists, then $[,]_{0}$ is a Lie bracket on $V$ and
$(V,[,]_{0})$ is called a \textit{contraction} of $(V,[,])$ \cite{Bur}. For
$0<\varepsilon\leq1$ the Lie algebras $(V,[,])$ are all isomorphic to
$(V,[,])$. Hence to obtain a new Lie algebra via contraction one needs
$\det\left(  f_{\varepsilon}\right)  =0$ for $\varepsilon=0$. This is a
necessary condition, but not a sufficient one.

We now present the contraction from the orthogonal group $SO(n+1)$ to the
Euclidean group $E(n)$. Moreover, we study the contraction from $S_{n}\sim
O(n+1)/O(n)$ to the Euclidean space $\mathbb{R}^{n}\sim E(n)/SO(n)$ in the
limit $R\rightarrow\infty$, where $R$ is the radius of $S_{n}.$We shall use
$R^{-1}$ as the contraction parameter. To realize the contraction explicitly,
let us introduce the Beltrami coordinates on the sphere $S_{n}$ of radius $R$
as $y_{\mu}=Ru_{\mu}/u_{0}$, $\mu=1,\ldots,n$, with
\begin{equation}
u_{0}^{2}+%
{\displaystyle\sum\limits_{k=1}^{n}}
u_{k}^{2}=R^{2},
\end{equation}
where $u_{i}$, $1\leq i\leq n$, are real coordinates. The sphere $S_{n}$
admits the group $SO(n+1)$ of isometries and can be realized as $S_{n}\sim
O(n+1)/O(n)$. Define the differential operators
\begin{equation}
L_{jk}=y_{j}\partial_{y_{k}}-y_{k}\partial_{y_{j}},\quad0\leq j,k,r,s\leq n,
\end{equation}
with the commutations relations%
\begin{equation}
\left[  L_{jk},L_{rs}\right]  =\delta_{kr}L_{js}+\delta_{js}L_{kr}-\delta
_{ks}L_{jr}-\delta_{jr}L_{ks},\quad0\leq j,k,r.s\leq n.
\end{equation}
The basis of the Lie algebra $so(n+1)$ of $SO(n+1)$ consists of $L_{jk}$,
$0\leq j<k\leq n$, and Laplace-Beltrami operator is given by
\begin{equation}
\Delta_{\mathrm{LB}}^{(n+1)}=\frac{1}{R^{2}}%
{\displaystyle\sum\limits_{0\leq j<k\leq n}^{n}}
L_{ij}.
\end{equation}
Introduce now the differential operators
\begin{equation}
q_{i}=R^{-2}L_{0i}=\partial_{y_{i}}+R^{-2}%
{\displaystyle\sum\limits_{k=1}^{n}}
y_{k}\partial_{y_{k}},\quad0\leq i\leq n,
\end{equation}%
\begin{equation}
\tilde{L}_{jk}=y_{j}q_{i}-y_{k}q_{i}=y_{j}\partial_{y_{k}}-y_{k}%
\partial_{y_{j}},\quad0\leq j,k\leq n,
\end{equation}
with the commutations relations%
\begin{equation}
\left[  \tilde{L}_{jk},\tilde{L}_{rs}\right]  =\delta_{kr}\tilde{L}%
_{js}+\delta_{js}\tilde{L}_{kr}-\delta_{ks}\tilde{L}_{jr}-\delta_{jr}\tilde
{L}_{ks},\quad1\leq j,k,r,s\leq n,
\end{equation}%
\begin{equation}
\left[  q_{i},\tilde{L}_{jk}\right]  =\delta_{ij}q_{k}-\delta_{ik}q_{j}%
,\quad\left[  q_{j},q_{k}\right]  =R^{-2}\tilde{L}_{jk},\quad1\leq i,j,k\leq
n.
\end{equation}
The Euclidean space $\mathbb{R}^{n}$ admits the group of isometries
$E(n)=\mathbb{R}^{n}\otimes SO(n)$, where the translation subgroup of the
semidirect product is identified to $\mathbb{R}^{n}$. Consider now the Killing
differential operators
\begin{equation}
p_{i}=x_{i,}\text{ }M_{jk}=x_{j}\partial_{x_{k}}-x_{k}\partial_{x_{j}}%
,\quad1\leq i,j,k\leq n,
\end{equation}
with the commutations relations%
\begin{equation}
\left[  p_{i},M_{jk}\right]  =\partial_{ij}p_{i}-\partial_{ik}p_{j}%
,\quad\left[  p_{i},p_{j}\right]  =0,
\end{equation}%
\begin{equation}
\left[  M_{jk},M_{rs}\right]  =\delta_{kr}L_{js}+\delta_{js}L_{kr}-\delta
_{ks}L_{jr}-\delta_{jr}L_{ks},\quad1\leq j,k,r,s\leq n.
\end{equation}
The basis of the Lie algebra $e(n)$ of $E(n)$ consists of $p_{i}$ and $M_{jk}
$, with $0\leq i\leq n$ and $0\leq j<k\leq n$, and Laplace-Beltrami operator
is given by
\begin{equation}
\Delta^{(n)}=%
{\displaystyle\sum\limits_{i=1}^{n}}
p_{i}^{2}.
\end{equation}
For $R\rightarrow\infty$, $SO(n+1)$ and $so(n+1)$ contract to $E(n)$ and
$e(n)$, respectively, and $y_{i}\rightarrow x_{i}$, $q_{i}\rightarrow p_{i}$,
$\tilde{L}_{jk}\rightarrow M_{jk}$, and $\Delta_{\mathrm{LB}}^{(n+1)}%
\rightarrow\Delta^{(n)}$.

When $\nu<-1$ , the zeros of $J_{\nu}$ are all real.%
\begin{equation}
J_{\nu}(z)=\left(  \frac{z}{2}\right)  ^{\nu}\lim_{n\rightarrow\infty}\frac
{1}{n^{\nu}}L_{n}^{\nu}\left(  \frac{z^{2}}{4n}\right)  ,\quad J_{\nu}%
(z)=\lim_{\lambda\rightarrow\infty}\lambda^{\nu}P_{\lambda}^{-\nu}\left(
\cos\frac{z}{\lambda}\right)  ,\label{lim}%
\end{equation}
where $L_{n}^{\nu}$ and $P_{\lambda}^{-\nu}$, are the Laguerre polynomial and the
associated Legendre polynomial, respectively \cite{BE}.

We now introduce the multipole Hamiltonian
\begin{equation}
H=H_{\mathrm{rad}}+\frac{f(\beta)}{\beta^{2}}\tilde{H}+V,
\end{equation}

\noindent where $H_{\mathrm{rad}}$ is the radial Hamiltonian, $\tilde{H}$ is
the angular Hamiltonian, $\beta=\sqrt{\sum\limits_{i=1}^{2n+1}x_{i}^{2}}$ is
the radial variable, and $f$ is an analytic function of $\beta^{2}$. In
particular: 1) $f(\beta)=1$ for the Bohr model \cite{Bohr} characterized by the group
chain $U(2n+1)\supset SO(2n+1)\supset SO(3)$; 2) $f(\beta)=(\beta/\beta
_{0})^{2}$, where $\beta_{0}$ is a constant parameter, for the models
$E(2n+1)$ and $X(2n+1)$ characterized by the contraction of the orthogonal
group $SO(2n+2)$ to the Euclidean group $E(2n+1)$. The Hamiltonian $H$ is
separable for $V=V_{\mathrm{rad}}+\tilde{V}f(\beta)/\beta^{2}$, where
$V_{\mathrm{rad}}$ is the radial potential Hamiltonian and $\tilde{V}$ is the
angular potential. If $V_{\mathrm{rad}}=C\beta^{-2}$, where $C$ is a constant
parameter, then the radial Schr\"{o}dinger equation is a differential equation
for the Bessel functions. The $E(2n+1)$ and $X(2n+1)$ models are characterized
by an infinite square-well potential in $\beta$ and the radial energies are
proportional to squared zeroes of Bessel functions. According to (\ref{lim}),
the zeros of the Bessel functions are approximated by the zeros of Laguerre
and associated Legendre polynomials. This approximation for the $E(5)$ model
is presented in \cite{RGF}. We obtain an analytic description for transitional
nuclei near critical points of quantum phase transitions \cite{CI}, \cite{CCI}.

\bigskip\ 

\section{\bigskip Semiclassical asymptotic}
\label{sec:level3}
\renewcommand{\theequation}{3.\arabic{equation}}
\setcounter{equation}{0}

In this section we study the semiclassical spectral properties of collective
Hamiltonians. Consider the following potential of nuclear surface quadrupole
oscillations of nuclei \cite{MG}:%
\begin{equation}
V_{n}=%
{\displaystyle\sum\limits_{\substack{m,m^{\prime}\geq0 \\2m+3m^{\prime}\leq n
}}}
c_{mm^{\prime}}\left(  q_{1}^{2}+q_{2}^{2}\right)  ^{m}\left(  q_{1}%
^{3}-3q_{1}q_{2}^{2}\right)  ^{m^{\prime}}.
\end{equation}
Let $q_{1}=\beta\cos\gamma$ and $q_{2}=\beta\sin\gamma$, where $\beta\geq0$
and $0\leq\gamma<2\pi$. Then%

\begin{equation}
V_{n}=%
{\displaystyle\sum\limits_{\substack{m,m^{\prime}\geq0 \\2m+3m^{\prime}\leq n
}}}
c_{mm^{\prime}}\beta^{2m+3m^{\prime}}\cos^{m^{\prime}}3\gamma.
\end{equation}
Consider the classical Hamiltonian \
\begin{equation}
\mathcal{H}_{n}=\frac{1}{2}\left(  p_{1}^{2}{+p}_{2}^{2}\right)  +V_{n},\quad
\end{equation}
where $n\geq2$. Then $\mathcal{H}_{3}$ is the Henon - Heiles Hamiltonian
\cite{HH}.

The critical points of $\mathcal{H}_{n}$ are given by $\partial V_{n}/\partial
q_{1}=\partial V_{n}/\partial q_{2}=0$. If $c_{10}>0$, then $q_{1}=q_{2}=0$ is
a minimum. The solutions of the system of polynomial equations
\begin{equation}
\frac{\partial V_{n}}{\partial q_{1}}=0,\quad\frac{\partial V_{n}}{\partial
q_{1}}=0,\quad\frac{\partial^{2}V_{n}}{\partial q_{1}^{2}}\frac{\partial
^{2}V_{n}}{\partial q_{2}^{2}}=\left(  \frac{\partial^{2}V_{n}}{\partial
q_{2}^{2}}\right)  ^{2}%
\end{equation}
form an algebraic set (separatrix) in the space of the control parameters
$c_{mm^{\prime}}$ and divides it into the regions where $V_{n}$ is
structurally stable. For $n=4$ consider $c_{10}$, $c_{20}>0$ and define
$\lambda=9c_{01}^{2}/(32c_{10}c_{20})$. If $0<$$\ \lambda<1$, then there
exists a minimum in the origin. The separatrix is given by $\lambda=1$. In the
region $\ \lambda>1$ there are four minima and three saddles.

\noindent We now define the Birkhoff-Gustavson normal forms \cite{Mos}. Let
$\mathbf{R}^{n}\times\mathbf{R}^{n}$ be the phase space endowed with the
canonical coordinates $(\mathbf{q},\mathbf{p})$, $\mathbf{q}=(q_{1}%
,\ldots,q_{n})$ and $\mathbf{p}=(p_{1},\ldots,p_{n})$. Let $K(q,p)$ be a
Hamiltonian function defined on a domain of $\mathbf{R}^{n}\times
\mathbf{R}^{n}$ centered in the origin $(\mathbf{0},\mathbf{0})$, which admits
the power-series expansion
\begin{equation}
K(\mathbf{q},\mathbf{p})=\sum_{j=1}^{n}\frac{\nu_{j}}{2}(p_{j}^{2}+q_{j}%
^{2})+\sum_{k=3}^{\infty}K_{k}(\mathbf{q},\mathbf{p}),
\end{equation}
around $(\mathbf{0},\mathbf{0})$, where each $K_{k}$ is a homogeneous
polynomial of degree $k$ in $(\mathbf{q},\mathbf{p})$, and $\nu_{j}$
non-vanishing constants. Let $(\mathbf{\xi},\mathbf{\eta})$, where
$\mathbf{\xi}=(\xi_{1},\ldots,\xi_{n})$ and $\mathbf{\eta}=(\eta_{1}%
,\ldots,\eta_{n})$, be another canonical coordinates of $\mathbf{R}^{n}%
\times\mathbf{R}^{n}$ and let the power series
\begin{equation}
W(\mathbf{q},\mathbf{\eta})=\sum_{j=1}^{n}q_{j}\eta_{j}+\sum_{k=3}^{\infty
}W_{k}(\mathbf{q},\mathbf{\eta})
\end{equation}
be a second type generating function, a function in the old position variables
$q$ and the new momentum variables $\eta$, associated with the canonical
transformation $(\mathbf{q},\mathbf{p})\rightarrow(\mathbf{\xi},\mathbf{\eta
})$, where $p_{j}=\partial S/\partial q_{j}$ and $\xi_{j}=\partial
S/\partial\eta_{j}$, $1\leq j\leq n$.\ Here each $W_{k}$ is a homogeneous
polynomial of degree $k$ in $(\mathbf{q},\mathbf{\eta})$. Then
\begin{equation}
G(\nabla_{\mathbf{\eta}}W,\mathbf{\eta})=H(\mathbf{q},\nabla_{\mathbf{q}}W),
\end{equation}

\begin{equation}
G(\mathbf{\xi},\mathbf{\eta})=\sum_{k\geq2}^{\infty}G_{k}(\mathbf{\xi
},\mathbf{\eta}),\quad G_{k}(\mathbf{\xi},\mathbf{\eta})=\sum_{j=1}^{n}%
\frac{\nu_{j}}{2}\left(  \eta_{j}^{2}+\xi_{j}^{2}\right)  ,\quad
\end{equation}
where each $G_{k}$ is a homogeneous polynomial of degree $k$ in $(\mathbf{\xi
},\mathbf{\eta})$.

The power series $G(\mathbf{\xi},\mathbf{\eta})$ is said to be in the
Birkhoff-Gustavson normal form up to degree $r$ if the Poisson commutation
relations $\left\{  G_{2},G_{k}\right\}  =0$ are satisfied for $k=3,\ldots,r$.
Here $\left\{  \cdot,\cdot\right\}  $ denotes the canonical Poisson bracket to
the coordinates $(\mathbf{\xi},\mathbf{\eta})$. We obtain the integrable form
\begin{equation}
K(I_{1},I_{2})=I_{1}+\frac{9c_{01}^{2}}{24}\left(  I_{1}^{2}-14I_{1}%
I_{2}+14I_{2}^{2}\right)  +\frac{c_{20}}{4}\left(  I_{1}^{2}+2I_{1}%
I_{2}-2I_{2}^{2}\right)  ,
\end{equation}
where $I_{1}$ and $I_{2}$ are action variables. The semiclassical quantum
spectrum is given by $K(n_{1}+1/2,n_{2}+1/2)$, where $n_{1}$ and $n_{2}$ are
non-negative integers.

Consider now the spherical multipole Hamiltonian%
\begin{equation}
H_{N}=\frac{1}{2}%
{\displaystyle\sum\limits_{i=1}^{2N+1}}
p_{i}^{2}+U,\quad U=%
{\displaystyle\sum\limits_{k\geq1}}
D_{k}\beta^{2k},\quad\beta=\sqrt{%
{\displaystyle\sum\limits_{i=1}^{2N+1}}
q_{i}^{2}}.
\end{equation}
The corresponding classical Hamiltonian is integrable. The radial
Schr\"{o}dinger equation can be written as
\begin{equation}
\left[  -{\frac{{\hbar}^{2}}{{2}}}\frac{\mathrm{d}^{2}}{\mathrm{d}{\beta}^{2}%
}+{\frac{{\hbar}^{2}l(l+N)}{2{\beta}^{2}}+U(\beta)}\right]  \Phi
=E\Phi.\label{Sch}%
\end{equation}
We apply the logarithmic perturbation theory via the $\hbar$-expansion
technique \cite{DT}. Using the substitution, $f({\beta})=\hbar\Phi^{-1}%
(d\Phi/d{\beta)}$, the Schr\"{o}dinger equation (\ref{Sch}) can be written as
a Riccati equation
\begin{equation}
\frac{{\hbar}}{{2}}\left(  \frac{\mathrm{d}f}{\mathrm{d}{\beta}}\right)
^{2}=\frac{{\hbar}^{2}l(l+N)}{2{\beta}^{2}}+{U(\beta)}-E.
\end{equation}
Consider the following series expansions in the Planck constant:
\begin{equation}
E=\sum_{n=0}^{\infty}{E_{n}\hbar^{n},}\;\;\;\;\;f({\beta})=\sum_{n=0}^{\infty
}f_{n}({\beta})\hbar^{n}.
\end{equation}
The quantization condition
\begin{equation}
\frac{1}{2\pi\mathrm{i}}\oint f{(r)\mathrm{d}\,\beta}=\hbar m,
\end{equation}
where $m$ is the number of zeros of the wave function inside the closed
contour, can be rewritten as%

\begin{equation}
\frac{1}{2\pi\mathrm{i}}\oint f{_{1}(\beta)\mathrm{d}\,\beta}=2n+l+N,\quad
\frac{1}{2\pi\imath}\oint f{_{n}(\beta)\mathrm{d}\,\beta}=0,\quad n>1,
\end{equation}
where $m=2n+l+N$. Here $n$ and $l$ are the radial and angular quantum numbers.
We obtain
\begin{align}
E_{1}  &  =\frac{1}{2}+m\,,\quad E_{2}=\frac{1}{2}\left(  3-2\,\lambda
+6{v}\right)  D{_{1}}\\
E_{3}  &  =\frac{1+2\,m}{8\,}\,\left[  \left(  -21+9\,\lambda-17\,{v}\right)
\,D{{_{1}}}^{2}+\,\left(  15-6\,\lambda+10\,{v}\right)  \,D{_{2}}\right]
\nonumber\\
E_{4}  &  =\frac{1}{128}[\left(  333-201\lambda+11\,\lambda^{2}-264\,\lambda
{v}+1041\,{v}+375\,{v}^{2}\right)  \,D_{1}^{3}\nonumber\\
&  -6\,\,\left(  60-39\lambda+3\lambda^{2}-42\,\lambda{v}+175{v}+55{v}%
^{2}\right)  \,D{_{1}}\,D{_{2}}\nonumber\\
&  +\,\left(  105-72\,\lambda+6\,\lambda^{2}-60\lambda\,{v}+\,\,280{v}%
+70{v}^{2}\right)  \,\,D{_{3}]}\nonumber
\end{align}
where ${v}=m\left(  m+1\right)  $ and $\lambda=l(l+N)$.

{\bf Acknowledgment} This paper was supported by the Romanian Ministry for Education and Research
under the grant CNCSIS, gr. A No 1069/2008.

\end{document}